\documentclass{article}
\usepackage{spconf,amsmath,graphicx}
\usepackage[colorlinks,linkcolor=blue]{hyperref}
\usepackage{caption}
\usepackage{multirow}
\usepackage{float}
\usepackage{setspace}
 
\usepackage{amsmath}
\usepackage[psamsfonts]{amssymb}
\title{Distance-based Weight Transfer for Fine-tuning from Near-field to Far-field Speaker Verification}
\name{Li Zhang$^1$, Qing Wang$^1$, Hongji Wang$^{2}$,
Yue Li$^1$, Wei Rao$^{2}$, Yannan Wang$^{2}$, Lei Xie$^{1,*}$\thanks{* Corresponding author.}} 
\address{
 $^1$Audio, Speech and Language Processing Group (ASLP@NPU), School of Computer Science, \\
Northwestern Polytechnical University (NPU), Xi'an, China  \\
  $^2$Tencent Ethereal Audio Lab, Tencent Corporation, Shenzhen, China
}
 
\begin{document}
\maketitle
 
\begin{abstract}
The scarcity of labeled far-field speech is a constraint for training superior far-field speaker verification systems. In general, fine-tuning the model pre-trained on large-scale near-field speech through a small amount of far-field speech substantially outperforms training from scratch. However, the vanilla fine-tuning suffers from two limitations~--~\textit{catastrophic forgetting} and \textit{overfitting}. In this paper, we propose a weight transfer regularization~(WTR) loss to constrain the distance of the weights between the pre-trained model and the fine-tuned model. With the WTR loss, the fine-tuning process takes advantage of the previously acquired discriminative ability from the large-scale near-field speech and avoids catastrophic forgetting.
Meanwhile, the analysis based on the PAC-Bayes generalization theory indicates that the WTR loss makes the fine-tuned model have a tighter generalization bound, thus mitigating the overfitting problem. Moreover, three different norm distances for weight transfer are explored, which are L1-norm distance, L2-norm distance, and Max-norm distance. We evaluate the effectiveness of the WTR loss on VoxCeleb~(pre-trained) and FFSVC~(fine-tuned) datasets. Experimental results show that the distance-based weight transfer fine-tuning strategy significantly outperforms vanilla fine-tuning and other competitive domain adaptation methods.
\end{abstract}
\section{Introduction}
\label{sec:intro}
Speaker verification~(SV) is a task to authenticate a speaker's identity given a small amount of speech from that speaker~\cite{naika2018overview}. In recent years, deep learning has shown remarkable success in SV tasks, but current methods often rely on a large amount of labeled training speech~\cite{nagrani2017voxceleb,desplanques2020ecapa}. The performance of SV systems degrades significantly in far-field conditions due to attenuated signals, noise interference as well as the rareness of far-field datasets~\cite{qin2020hi,movsner2022multisv}. In general, far-field speech datasets are relatively small in size and insufficient to train decent SV models from scratch. Therefore, near-field datasets are generally leveraged for training to improve the discriminative ability of SV systems~\cite{gusev2020stc,zhang2020npu}. However, there is a mismatch problem between near-field speech and far-field speech, so here transfer learning methods are necessary to transfer the SV model from near-field to far-field scenarios.

In recent SV research, there are four kinds of transfer learning methods typically used to address domain mismatch problems. 
The first one is domain adversarial training generally formulated as a min-max problem, where adversarial strategies~\cite{wang2018unsupervised,rohdin2019speaker} are used to confuse speaker encoders learning domain-invariant speaker representation. The second strategy is based on the back-end process of SV models. Unsupervised PLDA~\cite{garcia2014unsupervised} and CORAL+ PLDA~\cite{lee2019coral+} are proposed to adapt the between-class and within-class covariance matrices of PLDA models~\cite{prince2007probabilistic} in in-domain datasets. The third one is feature distribution alignment, including aligning the distribution between source and target domains~\cite{lin2020multi} and feature mapping with distance-based metric losses~\cite{movsner2019improving}. The last one is the simple fine-tuning strategy, which is a common and effective transfer learning method~\cite{gusev2020stc,zhang2020npu}.

In this paper, we mainly focus on fine-tuning strategy, which leverages large-scale near-field speech to improve the performance of SV systems in far-field scenarios. Compared with training from scratch, fine-tuning a pre-trained neural network using a far-field dataset can significantly improve
performance while reducing the far-field labeled speech requirements~\cite{gusev2020stc,zhang2020npu}. However, oracle fine-tuning just initializes
the weights of the target model with those of pre-trained
model without considering the catastrophic forgetting and
overfitting problems. 

To solve the above problems, we propose a weight transfer regularization~(WTR) loss to constrain the distance between the weights of the pre-trained model and those of the fine-tuned model. In addition, we analyze the generalization bound of fine-tuning with the WTR loss by PAC-Bayes theory~\cite{shawe1997pac}, which is proved to correlate better with empirical performance compared with several popular generalization theories~\cite{jiang2019fantastic}. 
The analysis result of the PAC-Bayes generalization bound testifies that the generalization bound is tighter with the WTR loss, which limits the distance between the
weights of the pre-trained and fine-tuned models than without any constraint. Furthermore, we explore three different norm distances in WTR loss. Experimental results on VoxCeleb and FFSVC datasets further demonstrate the effectiveness of WTR loss in fine-tuning.

\section{Weight Transfer for Fine-tuning}
\subsection{Weight Transfer Regularization}
The speaker verification framework mainly consists of two modules: speaker embedding extractor $f_E$ and classifier $f_C$. During fine-tuning, the learnable weights of the pre-trained speaker embedding extractor are used to initialize the soon-to-be-finetuned model in the first epoch. The classifier layer needs to be trained from scratch since different datasets contain different numbers of speakers. Suppose the learnable weights of the pre-trained and fine-tuned speaker embedding extractor are $W^s=[W_1^s, W_2^s, W_3^s, ..., W_j^s, ... , W_L^s]$ and $W^t=[W_1^t, W_2^t, W_3^t, ..., W_j^t, ... , W_L^t]$, where $L$ is the number of layers in the speaker embedding extractor. The data space of the large scale pre-training speech is $D_s=\{(x{^s_i},y{^s_i})\} \sim \mathcal{P}(s)$ consisting of $n_s$ labeled samples and the fine-tuning far-field dataset is $D_t=\{(x{^t_i}),(y{^t_i})\} \sim \mathcal{P}(t)$ with $n_t$ samples. In the fine-tuning process, only $D_t$ is available and $D_s$ is unavailable. 

The speaker prediction error ${L}^{(t)}\left(f_C,f_E\right)$ 
is measured by speaker classification loss AAMSoftmax~\cite{deng2019arcface} and the WTR loss in fine-tuning. The speaker embedding extractor in fine-tuning is initialized by the weights of pre-trained model. 
The total loss in fine-tuning the SV system is formulated as: 
\begin{equation} 
\begin{aligned}
 & \mathcal{L}^{(t)}\left(f_C,f_E\right)= \underset{(x_i^t, y_i^t) \sim \mathcal{P}(t)}{\mathbb{E}}\left[f_C\left(f_E(x_i^t), y_i^t\right)\right] \\
  &  + \alpha  \sum_{i=1}^{L} ||W_j^t - W^s ||_\pi, 
\end{aligned}
\label{feq1}
\end{equation}
where $\pi$ represents different norm distances and $\alpha$ is a trade-off hyper-parameter between the speaker classification loss and WTR loss. The $W^t_j$ denotes the learnable weights for the current $j$-th epoch during fine-tuning.

\subsection{Generalization Analysis of Weight Transfer}
To prove the WTR loss mitigating the overfitting of the fine-tuned model, we use the PAC-Bayes generalization theory~\cite{mcallester1999pac} to testify that limiting the distance between weights of the pre-trained and fine-tuned models can obtain a tighter generalization upper bound. 

The PAC-Bayes framework~\cite{mcallester1999pac} provides generalization guarantees for randomized inference predictors. Suppose the prior distribution of the pre-trained weights $W^t$ is $\mathcal{P}(s)$, which is independent of far-field speech datasets. The posterior distribution of the fine-tuned weights $W^t$ is $\mathcal{P}(s)$, which depends on the far-field training datasets. The PAC-Bayes theorem states that with probability at least $1-\delta$~($\delta \in (0,1)$) over 
the training data, the expected error of speaker classification can be bounded as follows~\cite{mcallester1999pac}:

\begin{equation}
\begin{aligned}
\underset{f \sim \mathcal{P}(t)}{\mathbb{E}}[\hat{\mathcal{L}(f)}] \leq \underset{f \sim \mathcal{P}(t)}{\mathbb{E}}[{\mathcal{L}}(f_C[f_E(x_i^t),y_i^t)])]+ \\ C 
\sqrt{\frac{\operatorname{KL}\left(\mathcal{P}(s) \| \mathcal{P}(t) \right)+3 \ln \frac{n}{\delta}+8}{n}},
\vspace{-2em}%
\end{aligned}
\label{seq2}
\end{equation}
where $C$ is the bound of the loss function and $n$ is the number of far-field samples.
Following the conclusions on the normal distribution of weights in convolutional neural networks from the previous work~\cite{huang2021rethinking}, we set the prior distribution $\mathcal{P}(s)=N(W^s, \delta^{2}Id)$, where $W^s$ is the weights
of the pre-trained network. The posterior distribution $\mathcal{P}(t)$ is centered at the fine-tuned model as
$N (W^t,\delta^2Id)$. 
We expand the KL divergence using the density of multivariate normal distributions as
\begin{equation}
\small
\begin{aligned}
\operatorname{KL}\left(\mathcal{P}(t)\|\mathcal{P}(s)\right) &=\underset{W_j^t \sim \mathcal{P}(t)}{\mathbb{E}}\left[\log \left(\frac{\operatorname{Pr}(W_j^t \sim \mathcal{P}(t))}{\operatorname{Pr}\left(W_j^t \sim \mathcal{P}(s)\right)}\right)\right] \\
&=\underset{W_j^t \sim \mathcal{P}(t)}{\mathbb{E}}\left[\log \frac{\exp \left(-\frac{1}{2 \sigma^2}\left\|W_j^t-W^s\right\|^2\right)}{\exp \left(-\frac{1}{2 \sigma^2}\|W_j^t-W^t\|^2\right)}\right] \\
&=-\frac{1}{2 \sigma^2} \underset{W_j^t \sim \mathcal{P}(t)}{\mathbb{E}}\left[\left\|W_j^t-W^s\right\|^2-\|W_j^t-W^t\|^2\right] \\
&=\frac{1}{2 \sigma^2} \underset{W_j^t \sim \mathcal{P}(t)}{\mathbb{E}}\left[\left\langle W^s-W^t, 2 W_j^t-W^s-W^t\right\rangle\right] \\
&=\frac{1}{2 \sigma^2}\left\|W^t- W^s\right\|_\pi^2, 
\end{aligned}
\label{teq3}
\end{equation}
\noindent where the $W^t_j$ denotes the learnable weights for the current $j$-th epoch during fine-tuning. In Eq.~\ref{seq2}, there are two variant items that decide the upper bound of generalization, which is the classification error and KL divergence of $\mathcal{P}(s)$ and $\mathcal{P}(s)$. The classification error is supervised by the classification loss AAMSoftmax~\cite{deng2019arcface} and speaker labels. By the proof of Eq.~\ref{teq3}, the magnitude of the KL divergence is positively related to the difference between the weights of the pre-trained and fine-tuned models. 

From the above proof, we can draw a conclusion that fine-tuning with WTR loss, which constrains the weight distance between the pre-trained model and the fine-tuned model, makes the fine-tuned model have a tighter generalization upper bound. In other words, fine-tuning with WTR loss mitigates the overfitting problem.

\subsection{Distance-based Weight Transfer}
To limit the weight difference between the pre-trained model and the fine-tuned model, we further explore three kinds of norm distance in WTR loss, which are L1-norm distance, L2-norm distance, and Max-norm distance, respectively.

\subsubsection{L1-norm Distance-based WTR}
The L1 norm distance is calculated as the sum of the absolute values of the difference between weights of pre-trained and fine-tuned models. 
The L1-norm-based WTR loss is formulated as:
\begin{equation}
||W_j^t - W^s ||_\pi = ||W_j^t - W^s||,
\label{l1norm}
\end{equation}
 where $W_j^t$ is the weight of the fine-tuning model on $j$\-th epoch and $W^s$ is the weight of the pre-trained model. 

\subsubsection{L2-norm Distance-based WTR}
The L2 norm is calculated as the square root of the sum of the squared of the difference between weights of the pre-trained and fine-tuned models. The L2-norm-based WTR loss is formulated as:
\begin{equation}
||W_j^t - W^s ||_\pi = ||W_j^t - W^s||_2^2,
\label{l2norm}
\end{equation}
 where $W_j^t$ is the weight of the fine-tuning model on $j$\-th epoch and $W^s$ is the weight of the pre-trained model. 
\subsubsection{Max-norm Distance-based WTR}
Max-norm distance is the largest of the absolute values of all elements in the difference matrix between weights of pre-trained and fine-tuned models. The Max-norm based WTR loss is formulated as:
\begin{equation}
\begin{aligned}
||W_j^t - W^s ||_\pi = \|{W_j^t - W^s}\|_{\infty} = \\ \max \left(\left\|(W_j^t - W^s)_1\right\|, \ldots,\left\|(W_j^t - W^s)_n\right\|\right),
\end{aligned}
\label{maxnorm}
\end{equation}
where $n$ denotes the columns of the $\|W_j^t - W^s)\|_1$. 
\section{Experimental Setup}
\subsection{Datasets}
We conduct experiments on VoxCeleb~(1\&2)~\cite{Nagrani19} and FFSVC ~\cite{qin2020ffsvc,qin2022far} datasets. VoxCeleb~(1\&2) is the large-scale pre-trained dataset. FFSVC~2020 and FFSVC~2022 are the two in-domain far-field datasets respectively. We test on two trials, which are the development trials of FFSVC~2022 and the development trials in task 2 of FFSVC~2020. Note we only use single-channel test utterances~(recorded by channel~2) in FFSVC~2020 trials. The development trials of FFSVC~2022 contain the utterances recorded by iPad and iPhone, so we select the iPhone and iPad recorded speech in FFSVC as the training set for the FFSVC~2022 trials. Meanwhile, we select the iPhone and channel~2 recorded data as the training set of the FFSVC~2020 trials.  
\subsection{Training Details}
In this paper, the structure of the speaker verification model is ECAPA-TDANN~(1024)~\cite{desplanques2020ecapa}. The loss function is additive angular margin softmax (AAM-softmax)~\cite{deng2019arcface} with a margin of 0.2 and a scale of 30. The speaker embedding models are trained with 80-dimensional log Mel-filter bank features with 25ms window size and 10ms window shift. In the pre-training process, the weight decay is set to 2e-5. The Adam optimizer with a cyclical learning rate varying between 1e-8 and 1e-3 following triangular policy~\cite{smith2017cyclical} is used for pre-training. The pre-trained model is trained for 12 epochs. Finally, total models are evaluated on the FFSVC~2020 and FFSVC~2022 test sets to find the best models.  

In the fine-tuning step, the neurons of the classification layer in the speaker verification model are modified to the speaker number of far-field speech and the learning rate are varying between 1e-8 to 1e-4. The other configurations are the same as the pre-training process. 

In the pre-training and fine-tuning steps, we all adopt online data augmentation, which includes adding noise~\cite{snyder2015musan}, adding reverberation~\cite{habets2006room} and specAug~\cite{park2019specaugment}. The hyperparameter of $\alpha$ in Eq.~\ref{feq1} is set to 0.01.  


 
\subsection{Comparison Methods}
We compare the WTR fine-tuning method with several other competitive domain adaptation methods. They are listed in the following:
\begin{itemize}
     \item Wasserstein Domain Adversarial Training~(Wasserstein DAT)~\cite{rohdin2019speaker}: The authors introduce an end-to-end domain adversarial method based on Wasserstein distance to mitigate the language mismatch problem in SV task.  
    \item Unsupervised PLDA~\cite{garcia2014unsupervised}: Daniel et.al use the out-of-domain PLDA system to cluster unlabeled in-domain speech, and then use the in-domain data to adapt the parameters of the PLDA system.
    \item CORAL+ PLDA~\cite{lee2019coral+}: Kong Aik et al. propose CORAL+ to compute the pseudo in-domain within and between class covariance matrices to regularize the corresponding matrices of PLDA. 
    \item MMD Transfer Learning~\cite{lin2020multi}: This work~\cite{lin2020multi} introduces a DNN-based adaptation method using maximum mean discrepancy~(MMD). 
\end{itemize}
When training the comparing method -- PLDA, we randomly select 1 million utterances in VoxCeleb~(1\&2) and the FFSVC dataset to train the initial PLDA, then only use the training dataset of FFSVC2020 or FFSVC2022 to train the adapted PLDA. Moreover, we reproduce the rest of the above methods on VoxCeleb~(1\&2), FFSVC~2020, and FFSVC~2022 datasets.

\subsection{ Scoring Criterion}
In the test phase, we use cosine similarity as the scoring criterion. The performance metrics are equal error rate (EER)
and minimum detection cost function (minDCF)~\cite{reynolds20172016} which is
evaluated with $P_{target} = 0.01$, $C_{miss} = C_{fa} = 1$.
\section{Experimental Results and Analysis}
\subsection{Results of Different Distance-based WTR}
The experimental results of different Distance-based WTR losses on FFSVC~2020 and FFSVC~2022 are listed in Table~\ref{table1}. From Table~\ref{table1}, we can observe that the EER/minDCF of the pre-trained model on FFSVC~2020 and FFSVC~2022 are 9.817\%/0.814 and 9.849\%/0.731. After vanilla fine-tuning, the EER/minDCF are reduced by 2.382\%/0.100 and 1.808\%/0.028 respectively, which illustrates that fine-tuning transfers the near-field model to the far-field speech to a certain extent. With the help of L2-norm distance-based WTR loss, the EER/minDCF are reduced by 1.548\%/0.041 and 1.399\%/0.143 compared with the results of vanilla fine-tuning on FFSVC~2020 and FFSVC~2022. As shown in Table~\ref{table1}, fine-tuning with L2-norm-based WTR loss obtains the lowest EER/minDCF on FFSVC~2020 and FFSVC~2022.
Each norm distance for WTR loss consistently outperforms the vanilla fine-tuning, demonstrating that the distance constraint between the weights of the pre-trained model and fine-tuned model not only keeps the transferability as vanilla fine-tuning but also alleviates the overfitting problem of the fine-tuned model.
\begin{table}[th]
\centering
\captionsetup{font={small}} 
\caption{EER/minDCF~(p=0.01) of fine-tuning with different distance-based WTR losses.}
\resizebox{\linewidth}{!}{
\begin{tabular}{ccccc}
\hline
\multirow{2}{*}{Method} & \multicolumn{2}{c}{FFSVC~2020} & \multicolumn{2}{c}{FFSVC~2022} \\ \cline{2-5} 
                        & EER~(\%)    & minDCF   & EER~(\%)    & minDCF   \\ \hline
Pre-trained Model        & 9.817      & 0.814            & 9.849      & 0.731            \\ \hline
Vanilla Fine-tuning        & 7.435      & 0.714            & 8.041      & 0.703            \\ \hline
+~WTR~(L1-norm)                 & 7.234      & 0.698            & 7.122      & 0.598            \\ \hline
+~WTR~(L2-norm)                 &\textbf{5.887}      & \textbf{0.673}            & \textbf{6.702}      & \textbf{0.560}            \\ \hline
+~WTR~(Max-norm)            & 6.478      & 0.698            & 7.088      & 0.615            \\ \hline
\end{tabular}
}
\label{table1}
\vspace{-2em}
\end{table}
\subsection{Analysis of Vanilla and WTR Fine-tuning}
\begin{figure}[th]
 \captionsetup{font={footnotesize}} 
\centering
\centerline{\includegraphics[width=\columnwidth]{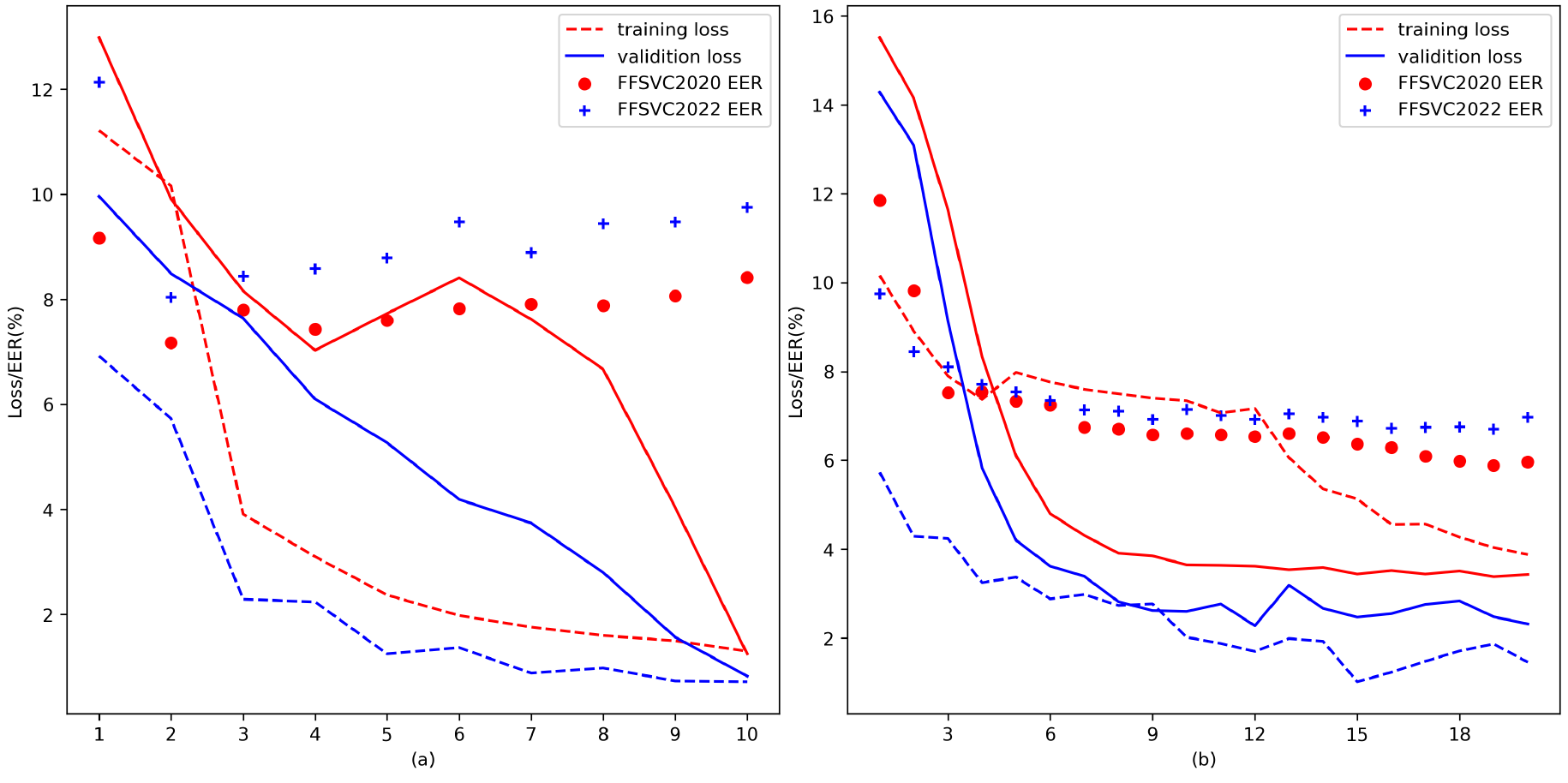}}
\caption{The trend of EER/loss values with the increasing of fine-tuning epochs. The red dashed/solid line is training/validation loss on FFSVC2020. The blue dashed/solid line is training/validation loss on FFSVC2022. (a)~vanilla fine-tuning, (b)~WTR fine-tuning.}
\label{fig:replace}
\vspace{-2em}%
\end{figure}

To show that weight transfer can avoid overfitting, we illustrate the loss and EER values during vanilla fine-tuning and fine-tuning with L2-norm-based WTR loss in Fig~\ref{fig:replace}. Fig~\ref{fig:replace}~(a) is the changing trend of the loss and EER values with the vanilla fine-tuning epochs. Although the loss function is decreasing~(the empirical error of the model is decreasing), the EER of the model is indeed increasing~(the generalization error keeps increasing). The results indicate that it is very easy to overfit during fine-tuning. In addition, the weights of the pre-trained speaker verification model are only used to initialize the fine-tuned model in the first epoch and there are no other constraints during fine-tuning, where the fine-tuned model is prone to forget the discriminability learned from large-scale near-field datasets. Compared with Fig~\ref{fig:replace}~(a), Fig~\ref{fig:replace}~(b) shows the tendency of the loss and EER values by the fine-tuning epochs with the help of WTR loss. Obviously, in Fig~\ref{fig:replace}~(b), the EER and loss of the training set and the validate set change in the same trend. Specifically, as the number of training epochs increases, the EER and loss get lower and lower until they stabilize at the 20th epoch. Therefore, the analysis further shows that WTR mitigates the overfitting of the model during fine-tuning.

\subsection{Comparison Results with Other Competitive Methods}
We compare the performance of WTR loss with other competitive domain adaptation methods in the SV task. The experimental results are shown in Table~\ref{table2}. We compare the WTR method with unsupervised PLDA adaptation, CORAL PLDA, Wasserstein DAT and MMD feature distribution alignment. As shown in Table~\ref{table2}, our proposed L2-norm distance-based WTR method outperforms all the compared domain adaptation methods.

\begin{table}[th]
\centering
\captionsetup{font={small}} 
\caption{EER/minDCF~(p=0.01) of other competitive methods.}
\resizebox{\linewidth}{!}{
\begin{tabular}{ccccc}
\hline
\multirow{2}{*}{Method} & \multicolumn{2}{c}{FFSVC~2020} & \multicolumn{2}{c}{FFSVC~2022} \\ \cline{2-5} 
                        & EER~(\%)    & minDCF   & EER~(\%)    & minDCF   \\ \hline
Unsuperivised PLDA~\cite{garcia2014unsupervised}       & 8.763     & 0.744            & 8.211     & 0.742            \\ \hline
CORAL+ PLDA~\cite{lee2019coral+}               & 7.435     & 0.714            & 7.837      & 0.724            \\ \hline
Wasserstein DAT~\cite{rohdin2019speaker}         &  8.433   &    0.778         & 9.136    & 0.715            \\ \hline
MMD~\cite{lin2020multi}                      & 7.335    & 0.725            & 7.503   & 0.619            \\ \hline
\textbf{WTR~(L2-norm)}     &\textbf{ 5.887}      & \textbf{0.673}            & \textbf{6.702}      & \textbf{0.560}             \\ \hline
\end{tabular}
}
\vspace{-2em}
\label{table2}
\end{table}

\section{Conclusion}
 In this paper, we propose a weight transfer regularization~(WTR) loss to solve the catastrophic forgetting and overfitting problems in fine-tuning far-field speaker verification models. Specifically, the WTR term is to limit the weight distance between the pre-trained model and the fine-tuned model. We also explore three kinds of norm distance in the WTR loss, which are L1-norm, L2-norm and Max-norm respectively. Moreover, we prove the generalization capacity of the fine-tuned model with WTR constraint by PAC-Bayes generalization theory. Experimental results and analysis on the FFSVC~2020 and FFSVC~2022 datasets demonstrate the effectiveness of the WTR term in alleviating overfitting and catastrophic forgetting problems during model fine-tuning.

\bibliographystyle{IEEE}
\begin{spacing}{0.98}  
\bibliography{refs}

\begin{thebibliography}{10}

\bibitem{naika2018overview}
Ravika Naika,
\newblock ``An overview of automatic speaker verification system,''
\newblock {\em Proc. ICICC}, pp. 603--610, 2018.

\bibitem{nagrani2017voxceleb}
Arsha Nagrani, Joon~Son Chung, and Andrew Zisserman,
\newblock ``Voxceleb: a large-scale speaker identification dataset,''
\newblock {\em Proc. INTERSPEECH}, 2017.

\bibitem{desplanques2020ecapa}
Brecht Desplanques, Jenthe Thienpondt, and Kris Demuynck,
\newblock ``{ECAPA-TDNN}: Emphasized channel attention, propagation and
  aggregation in tdnn based speaker verification,''
\newblock {\em INTERSPEECH}, 2020.

\bibitem{qin2020hi}
Xiaoyi Qin, Hui Bu, and Ming Li,
\newblock ``{Hi-Mia}: A far-field text-dependent speaker verification database
  and the baselines,''
\newblock in {\em ICASSP}. IEEE, 2020, pp. 7609--7613.

\bibitem{movsner2022multisv}
Ladislav Mo{\v{s}}ner, Old{\v{r}}ich Plchot, Luk{\'a}{\v{s}} Burget, and
  Jan~Honza {\v{C}}ernock{\`y},
\newblock ``Multisv: Dataset for far-field multi-channel speaker
  verification,''
\newblock in {\em ICASSP}. IEEE, 2022, pp. 7977--7981.

\bibitem{gusev2020stc}
Aleksei Gusev, Vladimir Volokhov, Alisa Vinogradova, Tseren Andzhukaev, Andrey
  Shulipa, Sergey Novoselov, Timur Pekhovsky, and Alexander Kozlov,
\newblock ``{STC}-innovation speaker recognition systems for far-field speaker
  verification challenge 2020.,''
\newblock in {\em INTERSPEECH}, 2020, pp. 3466--3470.

\bibitem{zhang2020npu}
Li~Zhang, Jian Wu, and Lei Xie,
\newblock ``{NPU} speaker verification system for interspeech 2020 far-field
  speaker verification challenge,''
\newblock {\em INTERSPEECH}, 2020.

\bibitem{wang2018unsupervised}
Qing Wang, Wei Rao, Sining Sun, Leib Xie, Eng~Siong Chng, and Haizhou Li,
\newblock ``Unsupervised domain adaptation via domain adversarial training for
  speaker recognition,''
\newblock in {\em ICASSP}. IEEE, 2018, pp. 4889--4893.

\bibitem{rohdin2019speaker}
Johan Rohdin, Themos Stafylakis, Anna Silnova, Hossein Zeinali, Luk{\'a}{\v{s}}
  Burget, and Old{\v{r}}ich Plchot,
\newblock ``Speaker verification using end-to-end adversarial language
  adaptation,''
\newblock in {\em ICASSP}. IEEE, 2019.

\bibitem{garcia2014unsupervised}
Daniel Garcia-Romero, Alan McCree, Stephen Shum, Niko Brummer, and Carlos
  Vaquero,
\newblock ``Unsupervised domain adaptation for i-vector speaker recognition,''
\newblock in {\em Odyssey}, 2014, vol.~8.

\bibitem{lee2019coral+}
Kong~Aik Lee, Qiongqiong Wang, and Takafumi Koshinaka,
\newblock ``The {CORAL+} algorithm for unsupervised domain adaptation of
  {PLDA},''
\newblock in {\em ICASSP}. IEEE, 2019.

\bibitem{prince2007probabilistic}
Simon~JD Prince and James~H Elder,
\newblock ``Probabilistic linear discriminant analysis for inferences about
  identity,''
\newblock in {\em ICCV}. IEEE, 2007, pp. 1--8.

\bibitem{lin2020multi}
Weiwei Lin, Man-Mai Mak, Na~Li, Dan Su, and Dong Yu,
\newblock ``Multi-level deep neural network adaptation for speaker verification
  using {MMD} and consistency regularization,''
\newblock in {\em ICASSP}. IEEE, 2020, pp. 6839--6843.

\bibitem{movsner2019improving}
Ladislav Mo{\v{s}}ner, Minhua Wu, Anirudh Raju, Sree Hari~Krishnan
  Parthasarathi, Kenichi Kumatani, Shiva Sundaram, Roland Maas, and Bj{\"o}rn
  Hoffmeister,
\newblock ``Improving noise robustness of automatic speech recognition via
  parallel data and teacher-student learning,''
\newblock in {\em ICASSP}. IEEE, 2019, pp. 6475--6479.

\bibitem{shawe1997pac}
John Shawe-Taylor and Robert~C Williamson,
\newblock ``A {PAC} analysis of a bayesian estimator,''
\newblock in {\em CLT}, 1997, pp. 2--9.

\bibitem{jiang2019fantastic}
Yiding Jiang, Behnam Neyshabur, Hossein Mobahi, Dilip Krishnan, and Samy
  Bengio,
\newblock ``Fantastic generalization measures and where to find them,''
\newblock {\em ICLR}, 2020.

\bibitem{deng2019arcface}
Jiankang Deng, Jia Guo, Niannan Xue, and Stefanos Zafeiriou,
\newblock ``Arcface: Additive angular margin loss for deep face recognition,''
\newblock in {\em CVPR}, 2019, pp. 4690--4699.

\bibitem{mcallester1999pac}
David~A McAllester,
\newblock ``{PAC-Bayesian} model averaging,''
\newblock in {\em {Computational Learning Theory}}, 1999, pp. 164--170.

\bibitem{huang2021rethinking}
Zhongzhan Huang, Wenqi Shao, Xinjiang Wang, Liang Lin, and Ping Luo,
\newblock ``Rethinking the pruning criteria for convolutional neural network,''
\newblock {\em Advances in Neural Information Processing Systems}, vol. 34, pp.
  16305--16318, 2021.

\bibitem{Nagrani19}
Arsha Nagrani, Joon~Son Chung, Weidi Xie, and Andrew Zisserman,
\newblock ``Voxceleb: Large-scale speaker verification in the wild,''
\newblock {\em Comput Speech Lang}, 2019.

\bibitem{qin2020ffsvc}
Xiaoyi Qin, Ming Li, Hui Bu, Rohan~Kumar Das, Wei Rao, Shrikanth Narayanan, and
  Haizhou Li,
\newblock ``The ffsvc 2020 evaluation plan,''
\newblock {\em INTERSPEECH}, 2020.

\bibitem{qin2022far}
Xiaoyi Qin, Ming Li, Hui Bu, Shrikanth Narayanan, and Haizhou Li,
\newblock ``Far-field speaker verification challenge (ffsvc) 2022: Challenge
  evaluation plan,'' 2022.

\bibitem{smith2017cyclical}
Leslie~N Smith,
\newblock ``Cyclical learning rates for training neural networks,''
\newblock in {\em WACV}. IEEE, 2017, pp. 464--472.

\bibitem{snyder2015musan}
David Snyder, Guoguo Chen, and Daniel Povey,
\newblock ``Musan: A music, speech, and noise corpus,''
\newblock {\em arXiv preprint arXiv:1510.08484}, 2015.

\bibitem{habets2006room}
Emanuel~AP Habets,
\newblock ``Room impulse response generator,''
\newblock {\em Technische Universiteit Eindhoven, Tech. Rep}.

\bibitem{park2019specaugment}
Daniel~S Park, William Chan, Yu~Zhang, Chung-Cheng Chiu, Barret Zoph, Ekin~D
  Cubuk, and Quoc~V Le,
\newblock ``{SpecAugment}: A simple data augmentation method for automatic
  speech recognition,''
\newblock {\em INTERSPEECH}, 2019.

\bibitem{reynolds20172016}
Douglas Reynolds, Elliot Singer, Seyed~O Sadjadi, Timothee Kheyrkhah, Audrey
  Tong, Craig Greenberg, Lisa Mason, and Jaime Hernandez-Cordero,
\newblock ``The 2016 {NIST} speaker recognition evaluation,''
\newblock Tech. {R}ep., MIT Lincoln Laboratory Lexington United States, 2017.

\end{thebibliography}
\end{spacing}
\end{document}